\documentclass[twocolumn,aps,superscriptaddress,amsmath,amssymb]{revtex4}
\usepackage{amsfonts}
\usepackage{bbm}

\usepackage{graphicx}
\usepackage{dcolumn}
\usepackage{bm}

\begin{document}

\title{Enhancement of electron transport and bandgap opening in graphene induced by adsorbates}
\author{Lan Chen}
\affiliation{School of Physics and Electronics, Hunan Key Laboratory for Super-microstructure and Ultrafast Process, Central South University, Changsha 410083, People's Republic of China}
\author{Fangping Ouyang}
\affiliation{School of Physics and Electronics, Hunan Key Laboratory for Super-microstructure and Ultrafast Process, Central South University, Changsha 410083, People's Republic of China}
\affiliation{Powder Metallurgy Research Institute, and State Key Laboratory of Powder Metallurgy, Central South University, Changsha 410083, People's Republic of China}
\affiliation{School of Physics and Technology, Xinjiang University, Urumqi 830046, People's Republic of China}
\author{Songshan Ma}
\affiliation{School of Physics and Electronics, Hunan Key Laboratory for Super-microstructure and Ultrafast Process, Central South University, Changsha 410083, People's Republic of China}
\author{Tie-Feng Fang}
\affiliation{School of Physical Science and Technology, Lanzhou University, Lanzhou 730000, China}
\author{Ai-Min Guo}
\email{aimin.guo@csu.edu.cn}
\affiliation{School of Physics and Electronics, Hunan Key Laboratory for Super-microstructure and Ultrafast Process, Central South University, Changsha 410083, People's Republic of China}
\author{Qing-Feng Sun}
\affiliation{International Center for Quantum Materials, School of Physics, Peking University, Beijing 100871, China}
\affiliation{Collaborative Innovation Center of Quantum Matter, Beijing 100871, China}
\affiliation{CAS Center for Excellence in Topological Quantum Computation, University of Chinese Academy of Sciences, Beijing 100190, China}

\begin{abstract}

Impurities are unavoidable during the preparation of graphene samples and play an important role in graphene's electronic properties when they are adsorbed on graphene surface. In this work, we study the electronic structures and transport properties of a two-terminal zigzag graphene nanoribbon (ZGNR) device whose scattering region is covered by various adsorbates within the framework of the tight-binding approximation, by taking into account the coupling strength $\gamma$ between adsorbates and carbon atoms, the adsorbate concentration $n_i$, and the on-site energy disorder of adsorbates. Our results indicate that when the scattering region is fully covered by homogeneous adsorbates, i.e., $n_i=1$, a transmission gap opens around the Dirac point and its width is almost proportional to $\gamma^2$. In particular, two conductance plateaus of $G=2e^2/h$ appear in the vicinity of the electron energy $E=\pm \gamma$. When the scattering region is partially covered by homogeneous adsorbates ($0<n_i<1$), the transmission gap still survives around the Dirac point even at low $n_i$, and its width is firstly increased by $n_i$ and then declined by further increasing $n_i$; whereas the conductance decreases with $n_i$ in the regime of low $n_i$ and increases with $n_i$ in the regime of high $n_i$. While in the presence of disordered adsorbates whose on-site energies are random variables characterized by the disorder degree, the transmission gap disappears at low $n_i$ and reappears at relatively high $n_i$. Furthermore, the transmission ability of the ZGNR device can be enhanced by the adsorbate disorder when the disorder degree surpasses a critical value, contrary to the localization picture that the conduction of a nanowire becomes poorer with increasing the disorder degree. The physics underlying these transport characteristics is discussed. Our results are in good agreement with experiments and may help for engineering graphene devices.

\end{abstract}

\maketitle

\section{Introduction\label{sec1}}

As the first two-dimensional material, graphene exhibits excellent physical properties \cite{dy1}, such as the Klein tunneling \cite{dy2,addsun2}, quantum spin Hall effect \cite{dy3,addsun1}, and high mobility of charge carriers \cite{dy4}. These properties promote the development of graphene in a wide variety of applications, such as transparent electrodes \cite{dy5}, sensors \cite{dy6,dy7}, thermally conductive composites \cite{dy8,dy9},
and electronic focusing devices \cite{xing1}. The scientific communities have devoted their great efforts to explore the physical properties of graphene \cite{neto1,sarma1,dy11,li1}. As demonstrated by the scanning tunneling microscopy \cite{dy12,dy13} and the atomic force microscope \cite{dy14,dy15}, it is very challenging to keep graphene surface atomically clean during preparation process and device fabrication process \cite{dy16}. In actual samples, various impurities coexist and are randomly adsorbed on graphene surface \cite{dy17}. The distribution and type of impurities depend on preparation method \cite{dy18}, environmental condition \cite{dy19}, and substrate \cite{dy20,dy21}. These impurities will interact with graphene and then modify its electronic structures and transport properties \cite{dy22}. The results of direct charge transport measurements on graphene are sometimes contradictory, indicating that it might be an insulator \cite{dy1,dy16,dy19,dy23} or a semiconductor \cite{dy2,dy3,dy4,dy5,dy24}. These different transport behaviors arise from a wide range of experimental complications, including the quality of graphene samples \cite{dy1,dy11,dy19}, gas atmosphere \cite{dy18}, and the interaction of graphene with substrate \cite{dy20,dy21,dy26}.

Several experiments showed that when adatoms, such as gold, tungsten, and indium, were deposited on graphene surface, they will supply electrons to graphene, which causes a strong scattering potential, reduces carrier mobility, and affects the output characteristics \cite{dy27,dy28,dy29,dy30}. According to the carrier scattering mechanism,  Trambly {\it et al.} classified adsorbates on graphene surface into two types, i.e., resonance scattering and non-resonance scattering \cite{dy31}. Brar {\it et al.} measured the gate-dependent $d{\rm I}/d{\rm V}$ spectra of graphene with cobalt adatoms, showing a bandgap of about 126 meV and an additional dip at 220 meV above the Fermi level \cite{dy32}. Besides, the effect of adsorbate concentration is also critical. A functionalized graphene decorated with oxygen-containing functional groups exhibits insulating behavior, where the bandgap width and the electrical conductivity can be tuned by reducing the oxygen content \cite{dy33}. Chen {\it et al.} studied the influence of potassium atoms at different deposition concentrations on the conductivity of graphene device as a function of gate voltage. They found that the width of the conductivity plateau increases with increasing concentration \cite{dy34}. Subsequent works found that Calcium atoms induce similar effects on the charge transport along graphene \cite{dy35}. Castellanos-Gomez {\it et al.} studied the electronic transport produced by the hydrogen adatom by means of the scanning tunneling spectroscopy, finding that hydrogen atoms induce a bandgap of about 0.4 eV \cite{dy36}. The bandgap opening of graphene, caused by hydrogen atoms, was also demonstrated by the angular resolved photoemission spectroscopy, and the width depends on hydrogen coverage \cite{dy37}. Recently, monolayer graphene synthesized by chemical vapor deposition with hydrogen plasma treatment has a reversible bandgap up to 3.9 eV \cite{dy38}.

Besides the experimental studies, many theoretical works were performed to understand the effects of adsorbates on the electronic properties of graphene. By employing the density functional theory, Wehling {\it et al.} studied the electronic properties of graphene adsorbed by various organic groups and a midgap state was found near the Dirac point \cite{dy41}. Robinson {\it et al.} presented a theory of electron transport along graphene with chemical adsorbates. They found that different types of adsorbates lead to the asymmetry of conductivity, which can be distinguished by p-type and n-type transport \cite{dy42}. Ihnatsenka {\it et al.} generalized the effective Hamiltonian of graphene with impurities and found that even at low adsorbate concentration the conductance is strongly suppressed and a transport gap develops near the Fermi energy \cite{dy43}. Yuan {\it et al.} studied the electronic structures of graphene with different adsorbate concentrations of hydrogen atoms. They showed that the adsorption of hydrogen atoms on graphene would affect its electrical and optical properties, and the bandgap can be generated when the coverage of hydrogen atoms was sufficiently large \cite{dy44}. Recently, Lee {\it et al.} performed theoretical calculations of the electronic properties by considering several adatoms, and revealed that these adatoms cause a specific bound state around the Dirac point and lead to unique spectral characteristics in the presence of a transverse magnetic field \cite{dy45}. In addition, previous theoretical works also focused on the possibility of adsorption sites \cite{dy46}, periodic structure \cite{dy47}, and spatial configuration \cite{dy48} of adsorbates on the transport properties of graphene. Based on the sensitivity of graphene to adsorbates, several groups designed gas-sensitive detectors \cite{dy7}, two-dimensional topological insulators \cite{dy49}, and valley filters \cite{dy50}.

\begin{figure}
\includegraphics[width=0.4\textwidth]{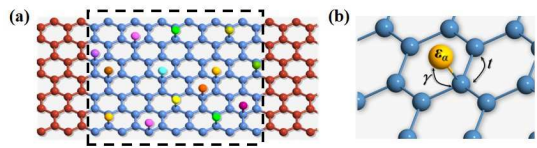}
\caption{\label{fig:1} (Color online) (a) Schematic view of a two-terminal ZGNR device composed of a scattering region (dashed rectangle) and two semi-infinite ZGNRs (wine carbon atoms) as electrodes. Here, the central scattering region is adsorbed by a variety of foreign impurities denoted by colorful balls. (b) Enlarged view of a single carbon atom connected to an impurity whose on-site energy is $\varepsilon_{\alpha}$ and coupling to the carbon atom is $\gamma.$}
\end{figure}

Although the effect of adsorbates has been extensively studied both experimentally and theoretically, there remain several issues need to be further clarified, such as the case of graphene surface covered by disordered adsorbates. In this paper, we report on a thorough study of the electronic structures and transport properties of a two-terminal zigzag graphene nanoribbon (ZGNR) whose scattering region is covered by various impurities, as illustrated in Fig.~\ref{fig:1}(a). We consider the influence of the coupling strength $\gamma$ between adsorbates and carbon atoms, the adsorbate concentration $n_i$, and the on-site energy disorder of adsorbates. Our results show that (i) when the scattering region is completely covered by a single type of adsorbates, a transmission gap appears around the Dirac point and its width is almost proportional to $\gamma^2$, leading to the insulating behavior of graphane \cite{dy23}. Although the transmission ability is declined by increasing $\gamma$, two transmission peaks develop in the vicinity of the Dirac point and their height is increased. In addition, two conductance plateaus quantized at $G=2e^2/h$ emerge around the electron energy $E=\pm \gamma$. (ii) When the scattering region is partially covered by homogeneous adsorbates, the transmission gap still persists at low $n_i$. In the regime of relatively low $n_i$, the gap width is increased by $n_i$ and the conductance is declined. While in the regime of high $n_i$, the gap width is decreased by further increasing $n_i$ and the conductance is enhanced. (iii) When the scattering region is partially covered by disordered adsorbates, the transmission gap is absent at low $n_i$ but reappears at high $n_i$. In particular, the conductance can be enhanced by the adsorbate disorder when the disorder degree surpasses a critical value.

The rest of the paper is organized as follows. In Sec.~\ref{sec2}, the model Hamiltonian and the numerical method are introduced. In Sec.~\ref{sec3}, the numerical results and discussion are presented. The results are concluded in Sec.~\ref{sec4}.

\section{Model and Method\label{sec2}}

The electron transport along a two-terminal ZGNR device whose scattering region is covered by various adsorbates, as illustrated in Fig.~\ref{fig:1}(a), can be simulated by the tight-binding Hamiltonian:
\begin{eqnarray}
{\cal H}=-t \sum_{\langle i, j\rangle}(c_{i}^{\dagger} c_{j}+ c_{j}^{\dagger} c_{i})+{\cal H}_{ad},\label{eq1}
\end{eqnarray}
where the first term describes a clean ZGNR system composed of a scattering region without any adsorbate and the left/right electrodes of two semi-infinite ZGNRs. $c_{i}^{\dagger}$ ($c_{i}$) is the creation (annihilation) operator at site $i$ of graphene lattice, and $t$ is the hopping integral between neighboring carbon atoms. The second term, representing adsorbates and their couplings to carbon atoms in the scattering region, is written as:
\begin{eqnarray}
{\cal H}_{ad}=\sum_{\alpha} \varepsilon_{\alpha} d_{\alpha}^{\dagger} d_{\alpha}+\gamma \sum_{\alpha} (d_{\alpha}^{\dagger} c_{p_\alpha} + c_{p_\alpha}^{\dagger}d_{\alpha} ).\label{eq2}
\end{eqnarray}
Here, $d_{\alpha}^{\dagger}$ ($d_{\alpha}$) is the creation (annihilation) operator at site $\alpha$ of adsorbates whose on-site energy is $\varepsilon_\alpha$, $p_\alpha$ is the site of carbon atoms coupled to adsorbates, and $\gamma$ is the hopping integral between carbon atoms and adsorbates, as shown in Fig.~\ref{fig:1}(b). By decimation of adsorbates, Eq.~(\ref{eq1}), describing a ZGNR device in the presence of various adsorbates, can be renormalized into the following form:
\begin{eqnarray}
{\cal H}=-t \sum_{\langle i, j\rangle} (c_{i}^{\dagger} c_{j} + c_{j}^{\dagger} c_{i} )+ \sum_{\alpha} V_{\alpha } c_{p_\alpha}^{\dagger } c_{p_\alpha},\label{eq3}
\end{eqnarray}
where the renormalized on-site energy is expressed as:
\begin{eqnarray}
V_{\alpha }= \frac{\gamma^ 2}{ E-\varepsilon_{\alpha}}, \label{eq4}
\end{eqnarray}
where $E$ is the electron energy. It is clear that the renormalized energy $V_\alpha$ is determined by three parameters, i.e., the electron energy $E$, the on-site energy $\varepsilon_{\alpha}$ of adsorbates, and the hopping integral $\gamma$. Notice that previous theoretical works usually focused on specific adsorbates, where both the parameters $\varepsilon_\alpha$ and $\gamma$ are fixed.

To understand the influence of adsorbates on the electronic structures and transport properties of the ZGNR device, the density of states is calculated and written as:
\begin{eqnarray}
\rho(E)=\sum_{k} \delta (E-E_{k} )=\operatorname{Tr}(\delta(E-{\cal H}_{ sc})), \label{eq6}
\end{eqnarray}
where ${\cal H}_{sc}$ is the Hamiltonian of the scattering region and $E_{k}$ is the corresponding eigenvalues. The density of states can be calculated by the kernel polynomial method which is based on the Chebyshev polynomials and has been widely used in other disordered systems \cite{dy55,dy56}.

The electron transport properties are calculated by using the scattering theory \cite{dy57}. For a two-terminal system, the transport modes are categorized into three types: incoming modes $\psi^{i}$, outgoing ones $\psi^{o}$, and evanescent ones $\psi^{e}$. The former two modes can propagate along the transport direction, whereas the last one decays quickly. Considering electrons transmitting along the $x$-aixs (the transport direction), the scattering states $\Psi^{el}$ in the left and right electrodes take the form \cite{dy58,dy59}:
\begin{eqnarray}
\Psi_{n}^{el}(x)=\psi_{n}^{i}(x)+\sum_{m=1}^{N_{o}} S_{m n} \psi_{m}^{o}(x)+\sum_{l=1}^{N_{e}} \tilde{S}_{l n} \psi_{l}^{e}(x), \label{eq7}
\end{eqnarray}
where $N_{o}$ ($N_{e}$) is the number of the outgoing (evanescent) modes. $S_{m n}$ ($\tilde{S}_{l n}$) is the scattering amplitude from an incoming mode $n$ to an outgoing mode $m$ (an evanescent mode $l$), which are the elements of the scattering matrix $S$. By matching the wavefunctions in the electrodes with those in the scattering region, i.e., $\Psi_{n}^{el}(x=x_b) =\Psi_n^S(x=x_b)$, one obtains the scattering matrix $S$, where $x_b$ is the border between the left/right electrode and the scattering region. Then, at zero temperature, the conductance can be calculated by employing the Landauer-B\"{u}ttiker formula \cite{dy57}:
\begin{eqnarray}
G=\frac{2 e^{2}}{h} T(E)=\frac{2 e^{2}}{h}\sum_{n\in L,m\in R}\left|S_{m n}\right|^{2}, \label{eq8}
\end{eqnarray}
where $L$ ($R$) denotes the left (right) electrode. In the present paper, all the numerical results are performed by employing the Kwant, a software package for quantum transport \cite{dy59}.

In the numerical results presented following, the width and length of the scattering region are set to $ D\approx10.5$ nm and $L\approx 51.2$ nm, respectively. We stress that the results still hold for the ZGNR devices with different sizes. The on-site energies of the carbon atoms are taken as the energy reference point, and the hopping integral $t$ between neighboring carbon atoms is set as the energy unit.

\section{Numerical Results and Discussion\label{sec3}}

\subsection{Electron transport along the ZGNR device fully covered by homogeneous adsorbates}

\begin{figure*}
\includegraphics[width=1.0\textwidth]{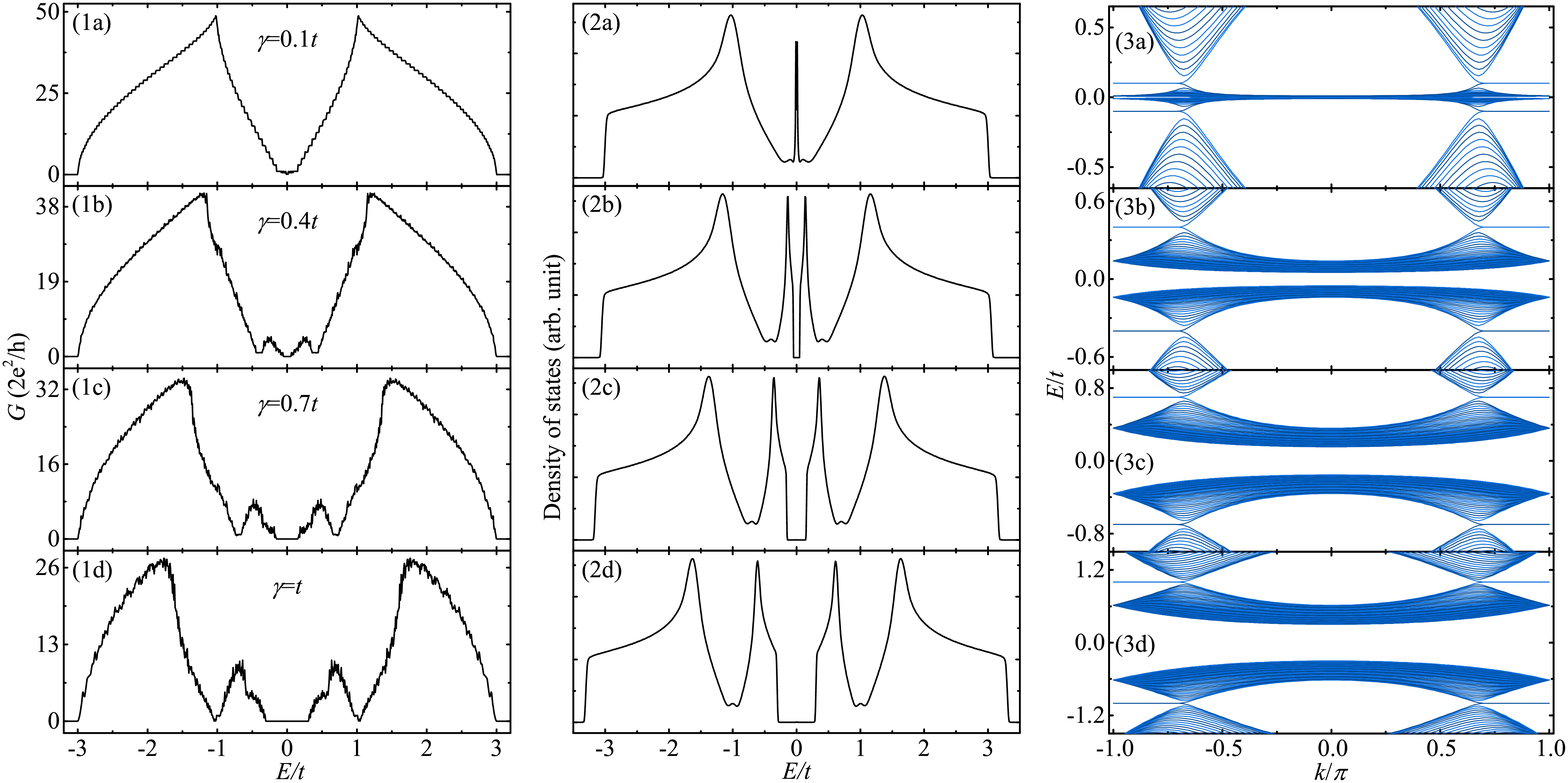}
\caption{\label{fig:2}(Color online) Electronic structure and transport properties of a ZGNR system whose scattering region is fully covered by identical adsorbates by considering various coupling strengths between carbon atoms and adsorbates. (1a-1d) Conductance $G$ vs electron energy $E$ of a two-terminal ZGNR device (left column), (2a-2d) density of states of the scattering region (middle column), and (3a-3d) dispersion relation of an infinite ZGNR completely covered by adsorbates with concentration $n_i=1$ (right column). The coupling strength between carbon atoms and adsorbates is chosen as $\gamma=0.1 t$ (top row), $\gamma=0.4 t$ (second row), $\gamma=0.7 t$ (third row), and $\gamma=t$ (bottom row).}
\end{figure*}

We first consider the simple case that the scattering region is fully covered by a single type of adsorbates with concentration being $n_i=1$, i.e., each carbon atom in the scattering region is connected to an identical adsorbate. This corresponds to totally functionalized graphene, such as completely hydrogenated graphene (graphane) \cite{dy23,dy36,dy38}, graphene oxide \cite{dy17,dy33}, and fluorinated graphene \cite{dy14,dy60}, which is regarded as novel functional materials. Besides, it might also pertain to the situation when graphene is epitaxially grown on a substrate, where all the carbon atoms interact with the substrate \cite{dy20}. Figures~\ref{fig:2}(1a-1d) show the conductance $G$ of a two-terminal ZGNR device as a function of electron energy $E$, while Figs.~\ref{fig:2}(2a-2d) plot the corresponding density of states of the scattering region, by taking into account various hopping integrals $\gamma$ between carbon atoms and adsorbates. The on-site energy of adsorbates is taken as $\varepsilon_\alpha=0$ for simplicity and similar results can be obtained when considering other values of $\varepsilon_\alpha$. Notice that each adsorbate can be treated as a single entity with specific potential energy and coupling to carbon atoms, as demonstrated by first-principles calculations \cite{dy41,dy43,dy61,dy62,dy63,dy64}. Besides, it was shown that the coupling between carbon atoms and adsorbates strongly depends on functionalization processes \cite{dy17,dy33}, and is very sensitive to the interaction between graphene and substrate \cite{dy20,dy22,dy64,dy65}.

\begin{figure}
\includegraphics[width=0.4\textwidth]{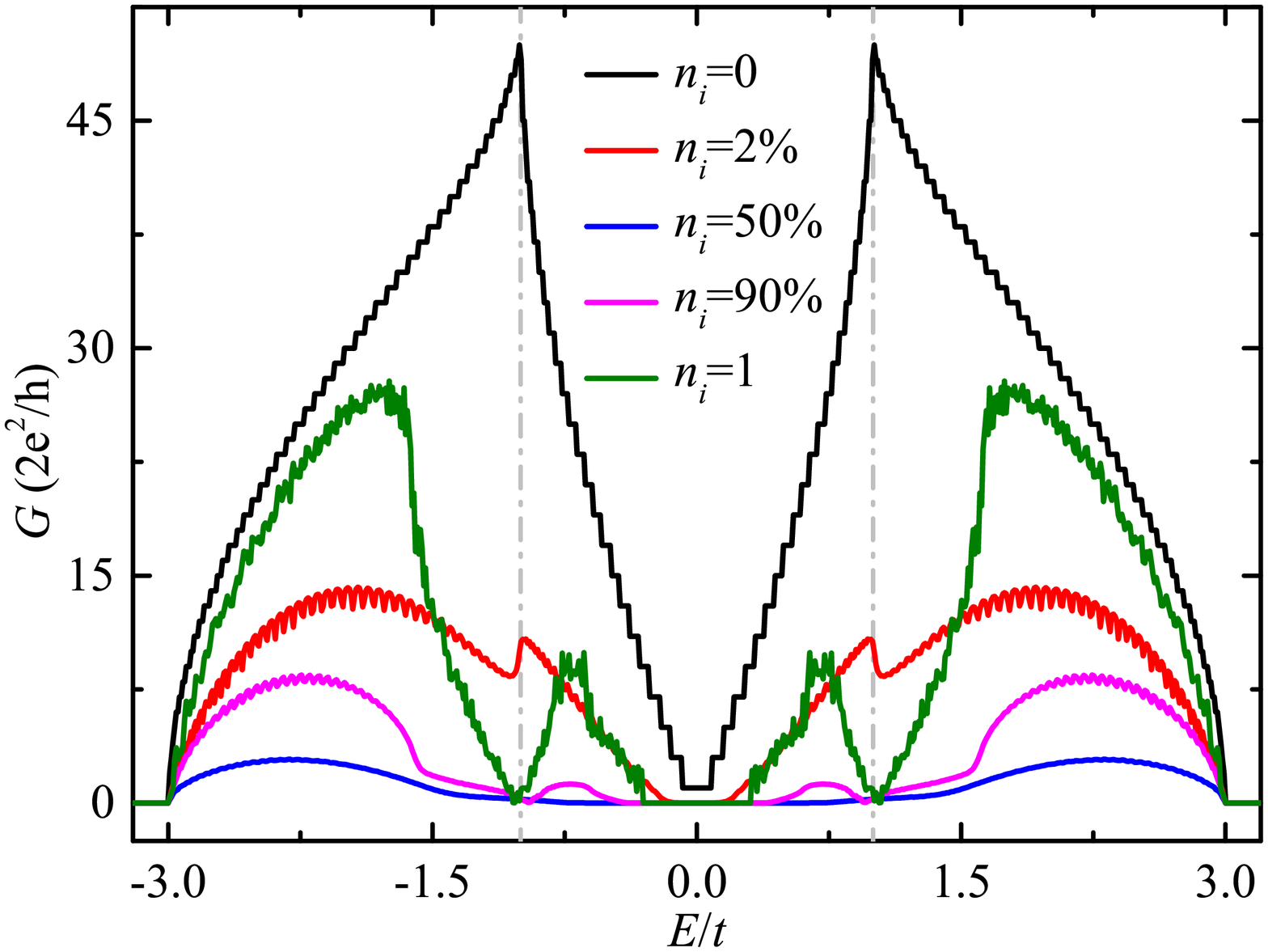}
\caption{\label{fig:3} (Color online) Electron transport along a ZGNR device with identical adsorbates randomly distributed on the scattering region. Conductance $G$ vs energy $E$ for typical values of adsorbate concentration, where the results are calculated from 2000 disorder configurations. The conductances of a ZGNR device without any adsorbate ($n_i=0$) and fully covered by adsorbates ($n_i=1$) are shown for reference, and the vertical dash-dotted line represents $E=t$.}
\end{figure}

As compared with the ideal case that a perfect conductance plateau of $G=2e^2/h$ exists around the Dirac point ($E=0$) for the ZGNR device in the absence of any adsorbate (see the black line in Fig.~\ref{fig:3}), it will be destroyed when \emph{}graphene is completely covered by homogeneous adsorbates, even for small $\gamma$. For instance, the conductance is declined, by about one order of magnitude, to $0.1e^2/h$ in the vicinity of the Dirac point when $\gamma=0.1t$ (Fig.~\ref{fig:2}(1a)). When the hopping integral $\gamma$ is enhanced and comparable to $t$, a transmission gap of $G=0$ can be clearly observed around the Dirac point. It is interesting to note that the width $E_{\text g}$ of this transmission gap is almost proportional to $\gamma^2$, i.e., $E_{\text g}\sim \gamma^2$, increasing from $E_{\text g} \approx 0.09t$ at $\gamma=0.4t$ to $E_{\text g} \approx 0.3t$ at $\gamma=0.7t$ and $E_{\text g} \approx 0.59t$ at $\gamma=t$. The emergence and enhancement of this bandgap, determined by adsorbates and their couplings to carbon atoms, can also be demonstrated in the density of states of the scattering region, as illustrated in Figs.~\ref{fig:2}(2a-2d). Further studies indicate that this phenomenon is almost independent of the system size, as clearly seen in Figs.~\ref{fig:2}(3a-3d) where the dispersion relation of an infinite ZGNR completely covered by adsorbates is displayed. This bandgap width $E_{\text g}$, extracted from different numerical results, is almost the same for a specific $\gamma$, implying that the adsorbate-induced bandgap opening is a general phenomenon. Our results are consistent with the first-principles calculations that the bandgap
increases with the interaction between graphene and substrate \cite{gg}, and qualitatively explain previous experiments \cite{dy20,dy23}. When graphene is physisorbed on SiC substrate, the interaction between graphene and substrate is small, and thus a relatively narrow bandgap is produced which is about 0.26 eV \cite{dy20}. While for completely hydrogenated graphene with $\gamma=2.2t$ \cite{dy42}, the bandgap is sharply increased to $E_{\text g} \approx 2.9t$. As a result, graphane behaves as an insulator, as evidenced by Raman spectroscopy and transmission electron microscopy \cite{dy23}.

This adsorbate-induced bandgap opening can be understood as follows. It can be inferred from Eq.~(\ref{eq4}) that when the electron energy is equal to the on-site energy $\varepsilon_ \alpha$ of adsorbates, the renormalized energy $V_ \alpha $ is infinite and the electron will be completely scattered at this site. This is the so-called antiresonant effect \cite{wangxr,gam}. In fact, this antiresonant effect can be extended to the following situation. When the electron energy locates within the range $[\varepsilon_\alpha -\Delta/2, \varepsilon_\alpha +\Delta/2]$ so that the renormalized energy satisfies $|V_\alpha| \gg E$, the electron can be strongly scattered at this site, leading to the localization phenomenon with conductance being zero. Although we cannot provide an analytic expression of the parameter $\Delta$ from Eq.~(\ref{eq4}), it is reasonable to assume that $\Delta$ should depend on $E$ and be proportional to $\gamma^2$. Since the on-site energy of adsorbates is $\varepsilon_\alpha =0$, the localization phenomenon, induced by the antiresonant-like effect, will occur when the electron energy locates within the range $[-\Delta/2, \Delta/2]$. As a result, a bandgap will appear around the Dirac point when the scattering region is fully covered by homogeneous adsorbates and its width is almost proportional to $\gamma^2$.

Besides the bandgap opening, one can see other interesting features. (i) The curve $G$-$E$ is symmetric with respect to the line $E=0$, independent of $\gamma$. This is due to the fact that the electron-hole symmetry is still preserved in the ZGNR device in the presence of identical adsorbates. (ii) Almost all the quantized conductance plateaus will be progressively destroyed by increasing $\gamma$ and the curve $G$-$E$ becomes rougher. However, we emphasize that two conductance plateaus of $G=2e^2/h$ still emerge around $E=\pm\gamma$, regardless of $\gamma$. When the electron energy approaches $\gamma$, the renormalized energy is approximated as $V_\alpha\sim E$ for $\varepsilon_ \alpha=0$. Then, the resonant tunneling mechanism dominates the electron transport process and the conductance plateaus appear around $E=\pm\gamma$, just as the one found around the Dirac point for the clean ZGNR device. In addition, two flat bands can be exactly observed at $E=\pm \gamma$, as can be seen in Figs.~\ref{fig:2}(3a-3d). (iii) The conductance is declined in general by increasing $\gamma$, because the renormalized sites serve as the potential barriers for $V_\alpha>E$ or the potential wells for $V_\alpha<E$ and their height/depth is monotonically increased. Nevertheless, when the hopping integral $\gamma$ is comparable to $t$, two additional transmission peaks emerge symmetrically near the Dirac point and their height is enhanced by increasing $\gamma$ (Figs.~\ref{fig:2}(1b-1d)), contrary to the decrement of the transmission ability discussed above. This originates from the appearance of additional conduction channels induced by adsorbates, as indicated by two sharp peaks near the Dirac point in the density of states (Figs.~\ref{fig:2}(2b-2d)), and these channels become more important for larger $\gamma$. Besides, these two peaks will be further separated from each other by increasing $\gamma$. In the following, we set $\gamma=t$ except for Fig.~\ref{fig:8}(c).

\begin{figure}
\includegraphics[width=0.4\textwidth]{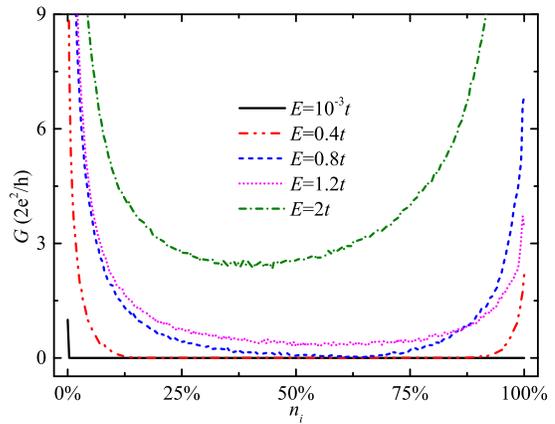}
\caption{\label{fig:4} (Color online) Electron transport along a ZGNR device with identical adsorbates randomly distributed on the scattering region. Conductance $G$ vs adsorbate concentration $n_i$ for typical values of electron energy $E$, where the results are calculated from 2000 disorder configurations.}
\end{figure}


\subsection{Electron transport along the ZGNR device partially covered by homogeneous adsorbates}

\begin{figure*}
\includegraphics[width=1.0\textwidth]{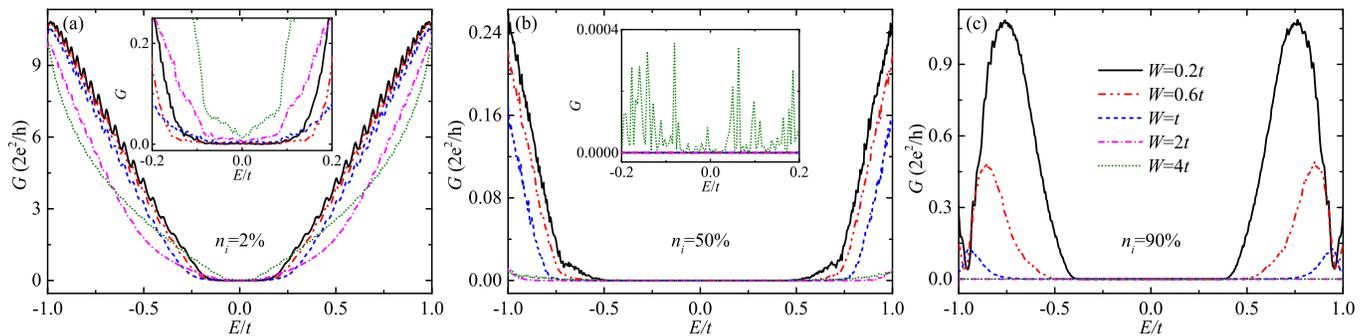}
\caption{\label{fig:6} (Color online) Electron transport along a ZGNR device with a variety of impurities randomly distributed on the scattering region. $G$ vs $E$ for typical values of adsorbate concentration $n_i$ and disorder degree $W$. (a) $n_{i}=2 \%$, (b) $n_{i}=50 \%$, and (c) $n_{i}=90 \%$, where the different curves denote the disorder degree of the on-site energies of the impurities. The inset shows the enlarged view of $G$-$E$ within the interval of $[-0.2t,0.2t]$.}
\end{figure*}

We then consider a more general case that the scattering region of the ZGNR device is partially covered by homogeneous adsorbates which are randomly distributed on the graphene surface. Figure~\ref{fig:3} plots the conductance $G$ for typical values of adsorbate concentration ranging from $n_i=0$ to $n_i=1$, as a function of electron energy $E$. Here, $n_i=0$ and $n_i=1$ are shown for reference, where the former corresponds to a ZGNR device in the absence of any adsorbate and the latter refers to the one fully covered by adsorbates. It is clear that the curve $G$-$E$ is also symmetric with respect to the line $E=0$, for whatever the values of $n_i$, because of the conservation of the electron-hole symmetry. In a clean ZGNR device, the transmission spectrum is characterized by many conductance plateaus quantized at integer multiples of $2e^2/h$ (see the black line in Fig.~\ref{fig:3}). However, these plateaus are fragile and can be destructed in the whole energy spectrum by partially covered adsorbates as well. When the electron energy is close to the Dirac point, a transmission gap of $G=0$ can be observed at small $n_i$ (see the red line in Fig.~\ref{fig:3}). This implies that full coverage is not a prerequisite ingredient to yield the bandgap opening in the graphene device. Instead, it can be achieved by a small amount of adsorbates randomly distributed on the scattering region, facilitating the bandgap engineering of the graphene device. Although the number of the potential barriers/wells produced by the adsorbates is increased by increasing $n_i$, the width $E_{\text g}$ of this transmission gap does not always increase with $n_i$. One notices that this gap width is increased from $E_{\text g} \approx 0.06t$ at $n_i=2\%$ to $E_{\text g} \approx 0.79t$ at $n_i=50\%$ and then decreased to $E_{\text g} \approx 0.66t$ at $n_i=90\%$ and $E_{\text g} \approx 0.59t$ at $n_i=1$. These results are consistent with previous experiments \cite{dy37,dy38,dh,fy,csp}. It was confirmed by several experimental groups that the bandgap of graphene increases with hydrogen coverage in the regime of relatively low $n_i$ \cite{dy37,dy38,dh}. Besides, other experiments demonstrated that the bandgap of graphene increases with humidity ratio when adsorbed by water molecules \cite{fy} and increases with the concentration of manganese oxide nanoparticles \cite{csp}.

Away from the Dirac point, the transmission ability does not decrease monotonically with increasing $n_i$. When the adsorbates initially adsorb on the clean ZGNR device, the conductance is sharply declined and exhibits oscillating behavior with increasing $E$ (see the red line in Fig.~\ref{fig:3}). For relatively low $n_i$, both the conductance and the oscillating amplitude are decreased by increasing $n_i$. While for large $n_i$, they are increased by increasing $n_i$ and the oscillating pattern of the fully covered ZGNR device becomes irregular (see the olive line in Fig.~\ref{fig:3}). This phenomenon can be further demonstrated in Fig.~\ref{fig:4}, where the conductance is displayed as a function of $n_i$ for several values of $E$. It can be seen from Fig.~\ref{fig:4} that the dependence of $G$ on $n_i$ is not monotonic. A turning point is found in the curves $G$-$n_i$ that the conductance decreases with $n_i$ for $n_i<n_c$ and then increases with $n_i$ for $n_i>n_c$.

The nonmonotonic dependence of $E_{\text g}$ on $n_i$ and $G$ on $n_i$ can be understood from the Anderson localization. In the absence of any adsorbate ($n_i=0$), the ZGNR device is a periodic system. When the adsorbates are initially introduced into the clean ZGNR device, the structural disorder appears and the system becomes disordered. Actually, the structural disorder strength is the largest at $n_i=50\%$ and the corresponding ZGNR device is the most disordered system, where half of the carbon atoms are randomly coupled to adsorbates. In the region of relatively low $n_i$, the structural disorder is enhanced by increasing $n_i$. As a result, the transmission gap becomes wider and the conductance is declined. In the region of large $n_i$, the structural disorder is decreased by further increasing $n_i$. Then, the transmission gap becomes narrower and the conductance is increased. When the scattering region is fully covered by adsorbates ($n_i=1$), the structural disorder vanishes and the ZGNR device can be regarded as an ordered system. In the case of extremely low (large) $n_i$ which is close to the order-disorder transition point, the electronic wave function is very sensitive to the structural disorder and the conductance will be dramatically changed by $n_i$. For example, when the electron energy locates around the Dirac point, the conductance quickly drops to zero for $n_i=1\%$ (see the black line in Fig.~\ref{fig:4}). This is consistent with previous studies that the transmission ability near the Dirac point is significantly affected by foreign impurities \cite{dy28,dy29,dy30,dy66}. However, the conductance tends to fluctuate around a certain value when the adsorbate concentration changes around $n_i=50\%$, because the electronic wave function is not sensitive to additional adsorbates when the structural disorder is sufficiently large.


\subsection{Electron transport along the ZGNR device partially covered by disordered adsorbates}

\begin{figure*}
\includegraphics[width=1.0\textwidth]{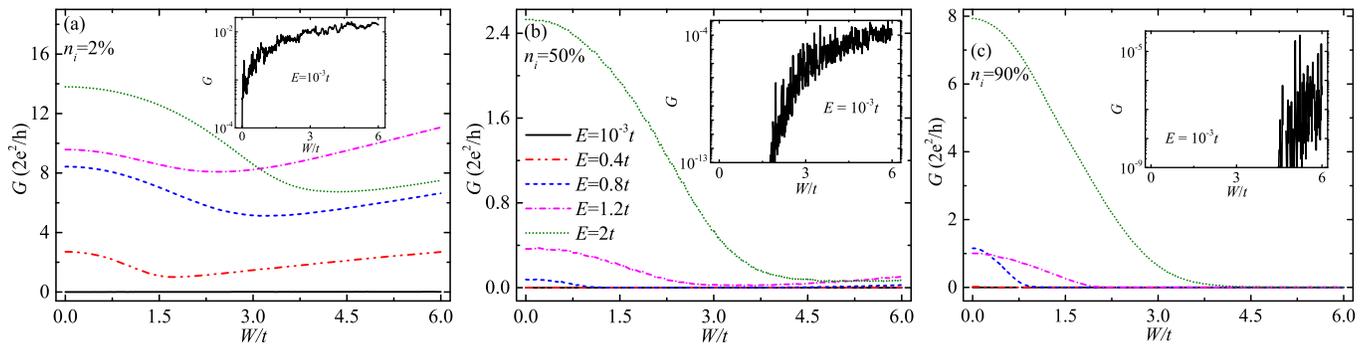}
\caption{\label{fig:7} (Color online) Electron transport along a ZGNR device with a variety of impurities randomly distributed on the scattering region. Conductance $G$ vs disorder degree $W$ for typical values of $E$ and $n_i$. (a) $n_{i}=2 \%$, (b) $n_{i}=50 \%$, and (c) $n_{i}=90 \%$, where different curves denote different electron energies $E$. The inset shows $\log G$-$W$ with $E=10^{-3}t$.}
\end{figure*}

Generally speaking, a variety of impurities may adsorb on graphene surface during the preparation procedure and interact with carbon atoms when the graphene is deposited on substrate. It would be suitable to simulate different experimental situations by choosing the on-site energies of adsorbates satisfying a certain disorder relationship. Here, we consider the most disordered situation that the scattering region is randomly covered by different adosrbates whose on-site energies $\varepsilon_\alpha$ are uniformly distributed within the range $[-W/2,W/2]$, with $W$ the disorder degree. Then, both the structural disorder and the on-site energy disorder coexist in the ZGNR device, leading to the emergence of several intriguing phenomena, as discussed below.

Figures~\ref{fig:6}(a)-\ref{fig:6}(c) plot the conductance $G$ for several values of $W$ and $n_i$, as a function of $E$. Here, different curves in each panel denote different disorder degrees $W$. Since the electron transport through the ZGNR device is mainly determined by the electrons whose energy is close to the Dirac point, the conductance is displayed within a small energy region $[-t,t]$. By inspecting Figs.~\ref{fig:6}(a)-\ref{fig:6}(c), one can identify several important features. (i) When the adsorbate concentration is small, the oscillating behavior of $G$ vs $E$ persists at small $W$ (see the black-solid line in Fig.~\ref{fig:6}(a)) and gradually disappears with increasing $W$ (see the olive-dotted line in Fig.~\ref{fig:6}(a)). It seems that the conductance beyond the region $[-0.66t,0.66t]$ is declined by increasing $W$. (ii) When the adsorbate concentration is increased up to $n_i=50\%$, the conductance is declined by about two orders of magnitude, because here the structural disorder strength is the largest. Meanwhile, the transmission gap develops and widens with increasing $W$, changing from $E_{\text g}=0.85t$ at $W=0.2t$ to $E_{\text g}=0.93t$ at $W=0.6t$, $E_{\text g}=1.04t$ at $W=t$, $E_{\text g}=1.19t$ at $W=2t$. (iii) When the adsorbate concentration is further increased up to $90\%$, the conductance is enhanced by approximately one order of magnitude as compared with the case of $n_i=50\%$, because of the decrement of the structural disorder. And two transmission peaks emerge in the energy spectrum and their height is declined by increasing $W$. Besides, the transmission gap becomes wider with increasing $W$, changing from $E_{\text g}=0.72t$ at $W=0.2t$ to $E_{\text g}=0.84t$ at $W=0.6t$, $E_{\text g}=1.1t$ at $W=t$, $E_{\text g}=1.93t$ at $W=2t$, and $E_{\text g}=2.1t$ at $W=4t$. It is clear that in the case of low $W$, $E_{\text g}$ at $n_i=50\%$ is larger than that at $n_i=90\%$, because the structural disorder dominates the electron transport property and is greater in the former case; while in the case of large $W$, $E_{\text g}$ at $n_i=50\%$ is smaller than that at $n_i=90\%$, because the number of the potential barriers/wells is less at $n_i=50\%$.

Besides, one can see other unusual phenomena. (i) At small $n_i$, the transmission gap vanishes in the presence of the on-site energy disorder and the conductance is increased by increasing $W$ (see the inset of Fig.~\ref{fig:6}(a)). (ii) At $n_i=50\%$, the transmission gap vanishes when the disorder degree is extremely large. To further understand the on-site energy disorder effect on the electron transport along the ZGNR device, Figs.~\ref{fig:7}(a)-\ref{fig:7}(c) present the conductance $G$ with typical values of $E$ and $n_i$, as a function of $W$. By inspecting Figs.~\ref{fig:7}(a)-\ref{fig:7}(c), a general trend of $G$ vs $W$ is observed for all the investigated $E$ and $n_i$. When the electron energy is away from the Dirac point, the dependence of $G$ on $W$ is not monotonic. A crossover is observed in all the curves $G$-$W$ that the conductance decreases with $W$ when $W<W_c$, whereas it is increased with increasing $W$ when $W>W_c$, especially in the case of low adsorbate concentration (Fig.~\ref{fig:7}(a)). This crossover $W_c$ is very sensitive to $E$.

\begin{figure*}
\includegraphics[width=1.0\textwidth]{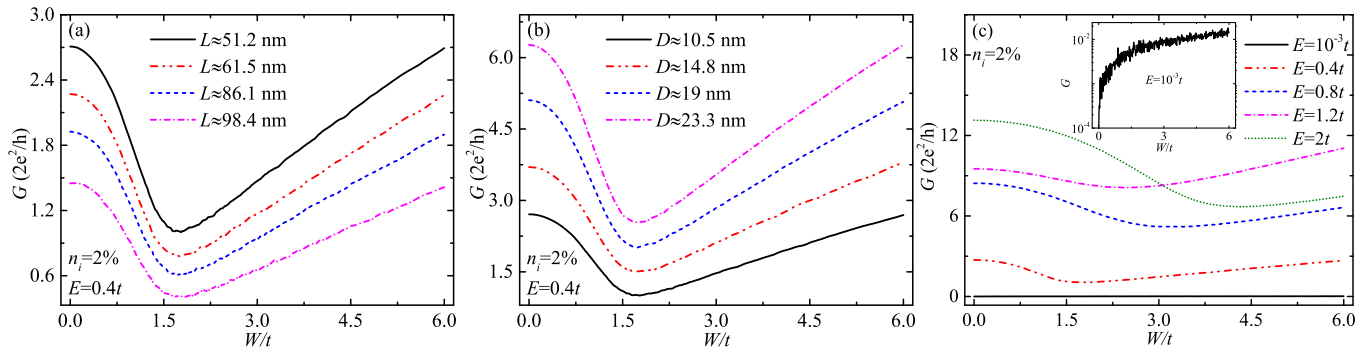}
\caption{\label{fig:8} (Color online) Electron transport along a ZGNR device with a variety of impurities randomly distributed on the scattering region by considering different device sizes and nonuniform coupling $\gamma$ between carbon atoms and adsorbates. Conductance $G$ vs disorder degree $W$ for (a) several values of length $L$ with $D\approx 10.5$ nm and (b) several values of width $D$ with $L\approx 51.2$ nm. The black-solid lines in (a) and (b) are shown for reference. (c) $G$ vs $W$ for typical values of $E$ by considering nonuniform $\gamma$ which is randomly distributed within the range $[0.8t, 1.2t]$. The inset shows $\log G$-$W$ with $E=10^{-3}t$. The other parameters are the same as Fig.~\ref{fig:7}(a).}
\end{figure*}

This can be understood from the antiresonant-like effect mentioned above. For an electron with energy $E$ transmitting through the ZGNR device, it will be dramatically scattered at carbon atoms connected to adsorbates whose on-site energy locates within the vicinity of $E$, i.e., $\varepsilon_ \alpha \in [E-\Delta /2, E+ \Delta/ 2]$. The closer $\varepsilon _\alpha$ to $E$, the stronger scattering the electron suffers and the stronger the antiresonant-like effect is. In the regime of relatively small $W$ where $\varepsilon _\alpha$ locates outside the vicinity of $E$, i.e., $W<2|E|$, the antiresonant-like effect will become stronger by increasing $W$ and consequently the transmission ability is decreased, consistent with the disorder-induced Anderson localization. While in the regime of relatively large $W$ where $\varepsilon _\alpha$ may locate within the vicinity of $E$, i.e., $W>2|E|$, the probability of adsorbates whose on-site energy locates within the vicinity of $E$ will be gradually declined by increasing $W$, i.e., the number of adsorbates exhibiting strong antiresonant-like effect will be decreased. Additionally, the on-site energy of adsorbates will be further away from $E$ and the corresponding antiresonant-like effect becomes weaker. As a result, the transmission ability will be enhanced by increasing $W$ when $W>2|E|$. In particular, when the electron energy is very close to the Dirac point, i.e., $E\sim0$, the antiresonant-like effect is progressively weakened by increasing $W$ and thus the conductance is increased, as illustrated in the insets of Fig.~\ref{fig:7}(a)-\ref{fig:7}(c). These results suggest that the impurity disorder-induced enhancement of transport may be a general phenomenon when the environment-induced disorder is sufficiently large; and one expects that the dependence of $W_c$ on $E$ may be approximated as $W\sim 2|E|$.

To demonstrate the robustness of the results, we investigate the electron transport along the ZGNR devices with different sizes and nonuniform coupling between carbon atoms and adsorbates, and take the case of $n_i=2\%$ as an example (Fig.~\ref{fig:7}(a)). Figures~\ref{fig:8}(a) and \ref{fig:8}(b) plot $G$ vs $W$ for several values of length $L$ and width $D$, respectively, with $n_i=2\%$ and $E=0.4t$. Here, the black-solid lines are shown for reference, which are identical to the red dash-dot-dot line in Fig.~\ref{fig:7}(a). It can be seen from Fig.~\ref{fig:8}(a) that the conductance is declined by increasing $L$ for whatever the values of $W$, because the electron suffers stronger scattering when transmitting along longer ZGNR devices. Contrarily, the conductance increases with the width $D$ (see Fig.~\ref{fig:8}(b)), because the number of conduction channels is enhanced by increasing $D$. Although the magnitude of the conductance is sensitive to the length and width, the nonmonotonic behavior of $G$ vs $W$ can still be observed for the ZGNR devices with different sizes. It is clear that the conductance decreases with $W$ for $W<W_c$ and increases with $W$ for $W>W_c$, for all the investigated length and width. In particular, the turning point $W_c$ is almost independent of the device sizes.

Note that the coupling $\gamma$ may differ from one type of adsorbates to another, we thus consider nonuniform $\gamma$ between carbon atoms and adsorbates. Here, the most disordered case is considered that $\gamma$ is randomly distributed within a certain range. Then, the scattering region contains three types of disorder, i.e., the structural disorder, the on-site energy one, and the coupling one. Figure~\ref{fig:8}(c) displays $G$ vs $W$ for typical values of $E$ with $n_i=2\%$ and $\gamma$ randomly distributed within $[0.8t, 1.2t]$. The only difference between Fig.~\ref{fig:8}(c) and Fig.~\ref{fig:7}(a) is that the coupling $\gamma$ is random in the former case and is uniform in the latter one. We find that although the conductance in Fig.~\ref{fig:8}(c) is different from that in Fig.~\ref{fig:7}(a), similar behavior can be observed in the curves $G$-$W$ when random coupling is taken into account. When the electron energy locates away from the Dirac point, the nonmonotonic behavior remains in the curves $G$-$W$ for whatever the values of $E$; when the electron energy is very close to the Dirac point, the conductance increases with $W$ (see the inset of Fig.~\ref{fig:8}(c)). Additionally, the turning point $W_c$ is nearly unchanged. Therefore, one can conclude that the results are robust and still hold for the ZGNR devices with different sizes and nonuniform coupling between carbon atoms and adsorbates.

\begin{figure}
\includegraphics[width=0.4\textwidth]{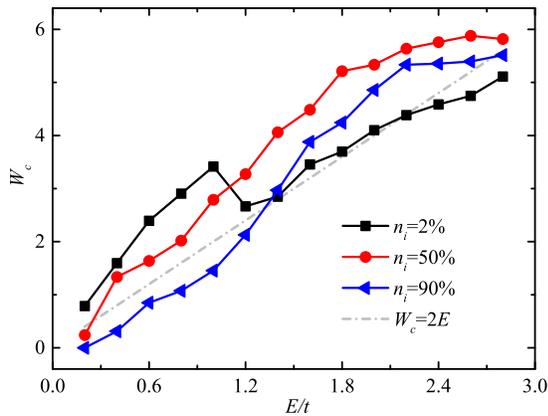}
\caption{\label{fig:9} (Color online) Turning point $W_c$ of the impurity disorder strength vs the electron energy $E$ for several adsorbate concentrations $n_i$ of a ZGNR device. Here, the dash-dotted line denotes the curve of $W_{c}=2E$.}
\end{figure}

Finally, Fig.~\ref{fig:9} displays the turning point $W_c$ of the adsorbate disorder strength with several $n_i$, as a function of $E$, which is extracted from Fig.~\ref{fig:7}. Here, the dash-dotted line represents the curve $W_c=2E$. For all the investigated $E$ and $n_i$, there always exists a turning point $W_c$, indicating that the adsorbate disorder-induced enhancement of transport is a generic feature for the ZGNR device when the disorder strength is sufficiently large. Besides, it is clear that the dependence of $W_c$ on $E$ follows the general trend of the curve $W_c=2E$ and increases with $E$, because the antiresonant-like effect dominates the electron transport along the ZGNR device. However, the curves $W_c$-$E$ deviate from each other for different $n_i$ and fluctuate along the curve $W_c=2E$, which is quite different from one-dimensional systems \cite{gam,gx}. In fact, besides the antiresonant-like effect, the electron transport is also affected by other physical mechanisms: (i) the structural disorder which is the largest at $n_i=50\%$; (ii) the number of potential barriers/wells which increases with increasing $n_i$; (iii) the on-site energy disorder of adsorbates. All these factors coexist and act simultaneously in the electron transport process, leading to the deviation of the curve $W_c$-$E$ from $W_c=2E$. In addition, since the parameter $\Delta$ is also determined by the electron energy, the deviation also depends on $E$.

\section{Conclusion\label{sec4}}

In summary, we investigate the electronic structures and transport properties of a zigzag graphene nanoribbon whose scattering region is adsorbed by various impurities. By using the scattering theory and the Landauer-Buttiker formula, the two-terminal conductance is calculated by taking into account the hopping integral $\gamma$ between carbon atoms and adsorbates, the adsorbate concentration $n_i$, and the on-site energy disorder of adsorbates. Our results indicate that a transmission gap develops around the Dirac point and its width is nearly proportional to $\gamma^2$ when the scattering region is fully covered by identical adsorbates, leading to the insulating behavior of graphane which is a completely hydrogenated graphene. Besides, two conductance plateaus are visible in the vicinity of $E=\pm \gamma$. This transmission gap still holds when the scattering region is partially covered by identical adsorbates. In the regime of low $n_i$, the width of the transmission gap increases with $n_i$ and the conductance decreases with $n_i$; while in the regime of high $n_i$, the width is declined by further increasing $n_i$ and the conductance is enhanced. When the scattering region is covered by disordered adsorbates, the transmission gap disappears at low $n_i$ and will reemerge by increasing $n_i$. In particular, the transmission ability of the graphene nanoribbon can be enhanced by the adsorbate disorder when the disorder strength of the on-site energies of adsorbates is sufficiently large, contrary to the localization picture that the conduction of a nanowire becomes poorer by increasing the disorder strength.

\section*{Acknowledgments}

L.C. thanks Shenglin Peng for implementation of the Kwant software package. This work is financially supported by the National Natural Science Foundation of China (No. 11504066, No. 11874187, No. 51272291, and No. 11921005), the Distinguished Young Scholar Foundation of Hunan Province (No. 2015JJ1020), the Innovation-Driven Project of Central South University (No. 2018CX044 and No. 2015CXS1035), and National Key R and D Program of China (Grant No. 2017YFA0303301).


\begin{references}

\bibitem{dy1}   A. K. Geim and K. S. Novoselov, Nat. Mater. {\bf 6}, 183 (2007).
\bibitem{dy2}	N. Stander, B. Huard, and D. Goldhaber-Gordon, Phys. Rev. Lett. {\bf 102}, 026807 (2009).
\bibitem{addsun2}
W. Long, Q.-F. Sun, and J. Wang, Phys. Rev. Lett. {\bf 101}, 166806 (2008).
\bibitem{dy3}	C. L. Kane and E. J. Mele, Phys. Rev. Lett. {\bf 95}, 226801 (2005).
\bibitem{addsun1}
Q.-F. Sun and X. C. Xie, Phys. Rev. Lett. {\bf 104}, 066805 (2010).
\bibitem{dy4}	P. Avouris, Nano Lett. {\bf 10}, 4285 (2010).
\bibitem{dy5}	S. Bae, S. J. Kim, D. Shin, J.-H. Ahn, and B. H. Hong, Phys. Scr. {\bf T146}, 014024 (2012).
\bibitem{dy6}	J. Duffy, J. Lawlor, C. Lewenkopf, and M. S. Ferreira, Phys. Rev. B {\bf  94}, 045417 (2016).
\bibitem{dy7}	F. Schedin, A. K. Geim, S. V. Morozov, E. W. Hill, P. Blake, M. I. Katsnelson, and K. S. Novoselov, Nat. Mater. {\bf  6}, 652 (2007).
\bibitem{dy8}	J. Liu, Q. Ma, Z. Huang, G. Liu, and H. Zhang, Adv. Mater. {\bf 31}, 1800696 (2019).
\bibitem{dy9}	A. C. Ferrari, F. Bonaccorso, and V. Fal¡¯ko, Nanoscale {\bf 7}, 4598 (2015).
\bibitem{xing1} Y. Xing, J. Wang, and Q.-F. Sun, Phys. Rev. B {\bf 81}, 165425 (2010).
\bibitem{neto1}	A. H. Castro Neto, F. Guinea, N. M. R. Peres, K. S. Novoselov, and A. K. Geim, Rev. Mod. Phys. {\bf 81}, 109 (2009).
\bibitem{sarma1}	S. Das Sarma, S. Adam, E. H. Hwang, and E. Rossi, Rev. Mod. Phys. {\bf 83}, 407 (2011).
\bibitem{dy11}	K. S. Novoselov, V. I. Fal'ko, L. Colombo, P. R. Gellert, M. G. Schwab, and K. Kim, Nature {\bf  490}, 192 (2012).
\bibitem{li1} G.-Y. Li, T.-F. Fang, A.-M. Guo, and Q.-F. Sun, Phys. Rev. B {\bf 100}, 115115 (2019).
\bibitem{dy12}	L. L. Patera, F. Bianchini, C. Africh, C. Dri, G. Soldano, M. M. Mariscal, M. Peressi, and G. Comelli, Science {\bf 359}, 1243 (2018).
\bibitem{dy13}	M. Narayanan Nair, M. Cranney, T. Jiang, S. Hajjar-Garreau, D. Aubel, F. Vonau, A. Florentin, E. Denys, M.-L. Bocquet, and L. Simon, Phys. Rev. B {\bf 94}, 075427 (2016).
\bibitem{dy14}	Y. Wang, W. C. Lee, K. K. Manga, P. K. Ang, J. Lu, Y. P. Liu, C. T. Lim, and K. P. Loh, Adv. Mater. {\bf  24}, 4285 (2012).
\bibitem{dy15}	S. Deng, D. Rhee, W.-K. Lee, S. Che, B. Keisham, V. Berry, and T. W. Odom, Nano Lett. {\bf 19}, 5640 (2019).
\bibitem{dy16}	G. Yang, L. Li, W. B. Lee, and M. C. Ng, Sci. Technol. Adv. Mater. {\bf 19}, 613 (2018).
\bibitem{dy17}	Q. Tang, Z. Zhou, and Z. Chen, Nanoscale {\bf 5}, 4541 (2013).
\bibitem{dy18}  B. Deng, Z. Liu, and H. Peng, Adv. Mater. {\bf 31}, 1800996 (2019).
\bibitem{dy19}	L. Lin, B. Deng, J. Sun, H. Peng, and Z. Liu, Chem. Rev.{\bf  118}, 9281 (2018).
\bibitem{dy20}	S. Y. Zhou, G.-H. Gweon, A. V. Fedorov, P. N. First, W. A. de Heer, D. H. Lee, F. Guinea, A. H. Castro Neto, and A. Lanzara, Nat. Mater. {\bf 6}, 770 (2007).
\bibitem{dy21}	Z. Hu, D. Prasad Sinha, J. Ung Lee, and M. Liehr, J. Appl. Phys. {\bf  115}, 194507 (2014).
\bibitem{dy22}	X. Xu, C. Liu, Z. Sun, T. Cao, Z. Zhang, E. Wang, Z. Liu, and K. Liu, Chem. Soc. Rev. {\bf  47}, (2018).
\bibitem{dy23}	D. C. Elias, R. R. Nair, T. M. G. Mohiuddin, S. V. Morozov, P. Blake, M. P. Halsall, A. C. Ferrari, D. W. Boukhvalov, M. I. Katsnelson, A. K. Geim, and K. S. Novoselov, Science {\bf 323}, 610 (2009).
\bibitem{dy24}  K. S. Novoselov, A. K. Geim, D. Jiang, and M. I. Katsnelson, Nature {\bf 438}, 197 (2005).
\bibitem{dy26}	N. Guo, K. M. Yam, and C. Zhang, npj 2D Materials and Applications {\bf 2},1 (2018).
\bibitem{dy27}	K. M. McCreary, K. Pi, A. G. Swartz, W. Han, W. Bao, C. N. Lau, F. Guinea, M. I. Katsnelson, and R. K. Kawakami, Phys. Rev. B {\bf 81}, 115453 (2010).
\bibitem{dy28}	J. A. Elias and E. A. Henriksen, Phys. Rev. B {\bf 95}, 075405 (2017).
\bibitem{dy29}	U. Chandni, E. A. Henriksen, and J. P. Eisenstein, Phys. Rev. B {\bf 91}, 245402 (2015).
\bibitem{dy30}	Z. Jia, B. Yan, J. Niu, Q. Han, R. Zhu, D. Yu, and X. Wu, Phys. Rev. B {\bf  91}, 085411 (2015).
\bibitem{dy31}	G. Trambly de Laissardiere and D. Mayou, Phys. Rev. Lett. {\bf  111}, 146601 (2013).
\bibitem{dy32}	V. W. Brar, R. Decker, H. Solowan, Y. Wang, L. Maserati, and K. T. Chan, Nat. Phys. {\bf 7}, 43 (2011).
\bibitem{dy33}	D. Chen, H. Feng, and J. Li, Chem. Rev. {\bf 112}, 6027 (2012).
\bibitem{dy34}	J.-H. Chen, C. Jang, S. Adam, M. S. Fuhrer, E. D. Williams, and M. Ishigami, Nat. Phys. {\bf 4}, 377 (2008).
\bibitem{dy35}	J. Katoch and M. Ishigami, Solid State Commun. {\bf  152}, 60 (2011).
\bibitem{dy36}	A. Castellanos-Gomez, M. Wojtaszek, Arramel, N. Tombros, and B. J. van Wees, Small {\bf  8}, 1607 (2012).
\bibitem{dy37}	R. Balog, B. J{\o}rgensen, L. Nilsson, M. Andersen, E. Rienks, M. Bianchi, M. Fanetti, E. L{\ae}gsgaard, A. Baraldi, S. Lizzit, Z. Sljivancanin, F. Besenbacher, B. Hammer, T. G. Pedersen, P. Hofmann, and L. Hornek{\ae}r, Nat. Mater. {\bf 9}, 315 (2010).
\bibitem{dy38}	J. Son, S. Lee, S. J. Kim, B. C. Park, H.-K. Lee, S. Kim, J. H. Kim, B. H. Hong, and J. Hong, Nat. Commun. {\bf  7}, 13261 (2016).
\bibitem{dy41}	T. O. Wehling, S. Yuan, A. I. Lichtenstein, A. K. Geim, and M. I. Katsnelson, Phys. Rev. Lett. {\bf 105}, 056802 (2010).
\bibitem{dy42}	J. P. Robinson, H. Schomerus, L. Oroszl\'{a}ny, and V. I. Fal'ko, Phys. Rev. Lett. {\bf  101}, 196803 (2008).
\bibitem{dy43}	S. Ihnatsenka and G. Kirczenow, Phys. Rev. B {\bf 83}, 245442 (2011).
\bibitem{dy44}	S. Yuan, H. De Raedt, and M. I. Katsnelson, Phys. Rev. B {\bf 82}, 115448 (2010).
\bibitem{dy45}	J. Lee, D. Kochan, and J. Fabian, Phys. Rev. B {\bf  99}, 035412 (2019).
\bibitem{dy46}	S. Irmer, D. Kochan, J. Lee, and J. Fabian, Phys. Rev. B {\bf 97}, 075417 (2018).
\bibitem{dy47}	A. Saffarzadeh and G. Kirczenow, Phys. Rev. B {\bf 90}, 155404 (2014).
\bibitem{dy48}	T. M. Radchenko, V. A. Tatarenko, I. Y. Sagalianov, Y. I. Prylutskyy, P. Szroeder, and S. Biniak, Carbon {\bf 101}, 37 (2016).
\bibitem{dy49}	J. Li, R.-L. Chu, J. K. Jain, and S.-Q. Shen, Phys. Rev. Lett. {\bf  102}, 136806 (2009).
\bibitem{dy50}	M. Azari and G. Kirczenow, Phys. Rev. B {\bf 97}, 245404 (2018).
\bibitem{dy55}	J. H. Pixley, P. Goswami, and S. Das Sarma, Phys. Rev. Lett. {\bf  115}, 076601 (2015).
\bibitem{dy56}	S. N. Taraskin and S. R. Elliott, Phys. Rev. B {\bf  65}, 052201 (2002).
\bibitem{dy57}	S. Datta, {\it Electronic Transport in Mesoscopic Systems} (Cambridge University Press, Cambridge, England, 1997).
\bibitem{dy58}	C. W. Groth, M. Wimmer, A. R. Akhmerov, and X. Waintal, New J. Phys. {\bf  16}, 063065 (2014).
\bibitem{dy59}	L. L. Li and F. M. Peeters, Phys. Rev. B  {\bf 97}, 075414 (2018).
\bibitem{dy60}	W. Feng, P. Long, Y. Feng, and Y. Li, Adv. Sci. {\bf  3}, 1500413 (2016).
\bibitem{dy61}	X. Liu, C. Z. Wang, Y. X. Yao, W. C. Lu, M. Hupalo, M. C. Tringides, and K. M. Ho, Phys. Rev. B {\bf 83}, 235411 (2011).
\bibitem{dy62}  S. Y. Davydov and G. I. Sabirova, Tech. Phys. Lett. {\bf 37}, 515 (2011).
\bibitem{dy63}  Kevin T. Chan, J. B. Neaton, and Marvin L. Cohen, Phys. Rev. B {\bf  77}, 235430 (2008).
\bibitem{dy64}  A. Gr\"{u}neis and Denis V. Vyalikh, Phys. Rev. B {\bf  77}, 193401 (2008).
\bibitem{dy65}  N. Kharche and V. Meunier, J. Phys. Chem. Lett. {\bf  7}, 1526 (2016).
\bibitem{gg}    G. Giovannetti, P. A. Khomyakov, G. Brocks, P. J. Kelly, and J. van den Brink, Phys. Rev. B {\bf 76}, 073103 (2007).
\bibitem{wangxr}X. R. Wang, Y. Wang, and Z. Z. Sun, Phys. Rev. B {\bf 65}, 193402 (2002).
\bibitem{gam}	A.-M. Guo, S.-J. Xiong, Z. Yang, and H.-J. Zhu, Phys. Rev. E {\bf 78}, 061922 (2008).
\bibitem{dh}	D. Haberer, D. V. Vyalikh, S. Taioli, B. Dora, M. Farjam, J. Fink, D. Marchenko, T. Pichler, K. Ziegler, S. Simonucci, M. S. Dresselhaus, M. Knupfer, B. B\"{u}chner, and A. Gr\"{u}neis, Nano Lett. {\bf 10}, 3360 (2010).
\bibitem{fy}	F. Yavari, C. Kritzinger, C. Gaire, L. Song, H. Gullapalli, T. Borca-Tasciuc, P. M. Ajayan, and N. Koratkar, Small {\bf 6}, 2535 (2010).
\bibitem{csp}	C. S. Park, Y. Zhao, J.-H. Lee, D. Whang, Y. Shon, Y.-H. Song, and C. J. Lee, Appl. Phys. Lett. {\bf 102}, 032106 (2013).
\bibitem{dy66}	K. Pi, W. Han, K. M. McCreary, A. G. Swartz, Y. Li, and R. K. Kawakami, Phys. Rev. Lett. {\bf 104}, 187201 (2010).
\bibitem{gx}	G. Xiong, Phys. Rev. B {\bf 76}, 153303 (2007).

\end{references}
\end{document}